\documentclass[11pt]{article}
\usepackage{times}
\usepackage{geometry}
\usepackage[utf8]{inputenc}
\usepackage{enumitem,amssymb}
\usepackage{graphicx}
\usepackage{fancyhdr}
\usepackage{aas_macros}
\usepackage{mdframed} 
\usepackage{wrapfig}
\usepackage{float}
\usepackage{multicol}
\usepackage{comment}
\usepackage[font=small,labelfont=bf]{caption}

\usepackage[authoryear]{natbib}
\setcitestyle{authoryear,open={(},close={)}}

\mdfdefinestyle{theoremstyle}{
innertopmargin=\topskip,}
\mdtheorem[style=theoremstyle]{lrptextbox}{}

\renewcommand{\thefootnote}{\fnsymbol{footnote}}

\usepackage{authblk}
\title{LRP 2020 Whitepaper: The Canadian Hydrogen Observatory and Radio-transient Detector (CHORD)}
\author[1,2]{K.~Vanderlinde \thanks{Corresponding Author: K. Vanderlinde, vanderlinde@dunlap.utoronto.ca}}
\author[3,4]{K.~Bandura}
\author[5]{L.~Belostotski}
\author[6]{R.~Bond}
\author[7,8]{P.~Boyle}
\author[10]{J.~Brown}
\author[7,8]{H.~C.~Chiang}
\author[7,8]{M.~Dobbs}
\author[1,2]{B.~Gaensler}
\author[11]{G.~Hinshaw}
\author[7,8]{V.~Kaspi}
\author[12]{T.~Landecker}
\author[7,8]{A.~Liu}
\author[13,14]{K.~Masui}
\author[13]{J.~Mena-Parra}
\author[2]{C.~Ng}
\author[6]{U.~Pen}
\author[12]{M.~Rupen}
\author[7,8,15]{J.~Sievers}
\author[16]{K.~Smith}
\author[17]{K.~Spekkens}
\author[6]{I.~Stairs}
\author[16]{N.~Turok}

\affil[1]{Department of Astronomy \& Astrophysics, University of Toronto, 50 St George St, Toronto, ON, M5S 3H4, Canada}
\affil[2]{Dunlap Institute for Astronomy \& Astrophysics, University of Toronto, 50 St George St, Toronto, ON, M5S 3H4, Canada}
\affil[3]{CSEE, West Virginia University, Morgantown, WV 26505, USA}
\affil[4]{Center for Gravitational Waves and Cosmology, West Virginia University, Morgantown, WV 26505, USA}
\affil[5]{Department of Electrical and Computer Engineering, University of Calgary, Calgary, AB T2N 1N4, Canada}
\affil[6]{Canadian Institute for Theoretical Astrophysics, University of Toronto, 60 St. George Street, Toronto, ON M5S 3H8, Canada}
\affil[7]{Department of Physics, McGill University, 3600 rue University, Montr\'eal, QC H3A 2T8, Canada}
\affil[8]{McGill Space Institute, McGill University, 3550 rue University, Montr\'eal, QC H3A 2A7, Canada}
\affil[9]{School of Mathematics, Statistics \& Computer Science, University of KwaZulu-Natal, Private Bag X54001, Durban 4000, South Africa}
\affil[10]{Department of Physics and Astronomy, University of Calgary, Canada, T2N 1N4}
\affil[11]{Department of Physics and Astronomy, University of British Columbia, 6224 Agricultural Road, Vancouver, BC V6T 1Z1, Canada}
\affil[12]{Dominion Radio Astrophysical Observatory, Herzberg Astronomy \& Astrophysics Research Centre, National Reseach Council Canada, P.O. Box 248, Penticton, BC V2A 6J9, Canada}
\affil[13]{MIT Kavli Institute for Astrophysics and Space Research, Massachusetts Institute of Technology, 77 Massachusetts Avenue, Cambridge, MA 02139, United States}
\affil[14]{Department of Physics, Massachusetts Institute of Technology, 77 Massachusetts Avenue, Cambridge, MA 02139, United States}
\affil[15]{School of Chemistry and Physics, University of KwaZulu-Natal, Private Bag X54001, Durban 4000, South Africa}
\affil[16]{Perimeter Institute for Theoretical Astrophysics, 31 Caroline Street N, Waterloo, ON N2L 2Y5, Canada}
\affil[17]{Department of Physics and Space Sciences, Royal Military College of Canada, PO Box 17000, Station Forces, Kingston, ON K7K 7B4, Canada}
\date{}

\begin{document}
% ****BEGIN MAIN WHITE PAPER SECTION****

% Instructions: Please insert your white paper text here.
%A white paper should be a self-contained description of a future opportunity for Canadian astronomy. A white paper will be most effective and useful if it concisely summarises and recommends an option that the LRP2020 panel should be considering for prioritisation.
%
% White papers are not required to contain a specific set of sections or headings. Depending on the content, the following topics may be appropriate to include:
%connection or relevance to Canada
%timeline 
%cost 
%description of risk
%governance / membership structure 
%justification for private submission of supplementary information

% The full document can have a maximum length of 10 pages including text, figures, tables, responses to selection criteria, references and appendices, and a size of 30 MB.

%%%%%%%%%%%%%%%

\clearpage
\maketitle
\thispagestyle{empty}

%%%%%%%%%%%%%%%%%%%%%%%%%%%%%
%%%%%%%%%%%%%%%%%%%%%%%%%%%%%
%%%   START OF ABSTRACT   %%%
%%%%%%%%%%%%%%%%%%%%%%%%%%%%%
%%%%%%%%%%%%%%%%%%%%%%%%%%%%%
\begin{abstract}
The Canadian Hydrogen Observatory and Radio-transient Detector (CHORD) is a next-generation radio telescope, proposed for construction to start immediately.
%, which will leverage Canadian technology developments to yield breakthrough measurements of the cosmos.
CHORD is a pan-Canadian project, designed to work with and build on the success of the Canadian Hydrogen Intensity Mapping Experiment (CHIME). It is an ultra-wideband, ``large-N, small-D'' telescope, consisting of a central array of 512$\times$ 6-m dishes, supported by a pair of distant outrigger stations, each equipped with CHIME-like cylinders and a 64-dish array.
%With breakthrough sensitivity, bandwidth, and localization capabilities,
CHORD will measure the distribution of matter over a huge swath of the Universe, detect and localize tens of thousands of Fast Radio Bursts (FRBs), and undertake cutting-edge measurements of fundamental physics.
\end{abstract}
%%%%%%%%%%%%%%%%%%%%%%%%%%%%%
%%%%%%%%%%%%%%%%%%%%%%%%%%%%%
%%%    END OF ABSTRACT    %%%
%%%%%%%%%%%%%%%%%%%%%%%%%%%%%
%%%%%%%%%%%%%%%%%%%%%%%%%%%%%
\newpage

%\tableofcontents
\setcounter{page}{1}

\section*{Introduction}
Radio astronomy has undergone a renaissance in recent years, driven by technological and scientific advances sweeping society.
The incredible expansion of available computing power has opened new doors for instrumental design. The explosive growth of telecommunications has fueled a host of critical front-end technologies, and novel construction materials and techniques have paved new avenues for telescope design.
%The advancement of Intensity Mapping as a way to survey the Universe, advances in fundamental physics, and the discovery of Fast Radio Bursts, have all thrown open whole new fields to exploration.
Advances across the fields of radio astronomy, time-domain astrophysics, and cosmology have opened new fields to exploration.

\begin{figure}[H]
\centering
    \includegraphics[width=0.9\linewidth]{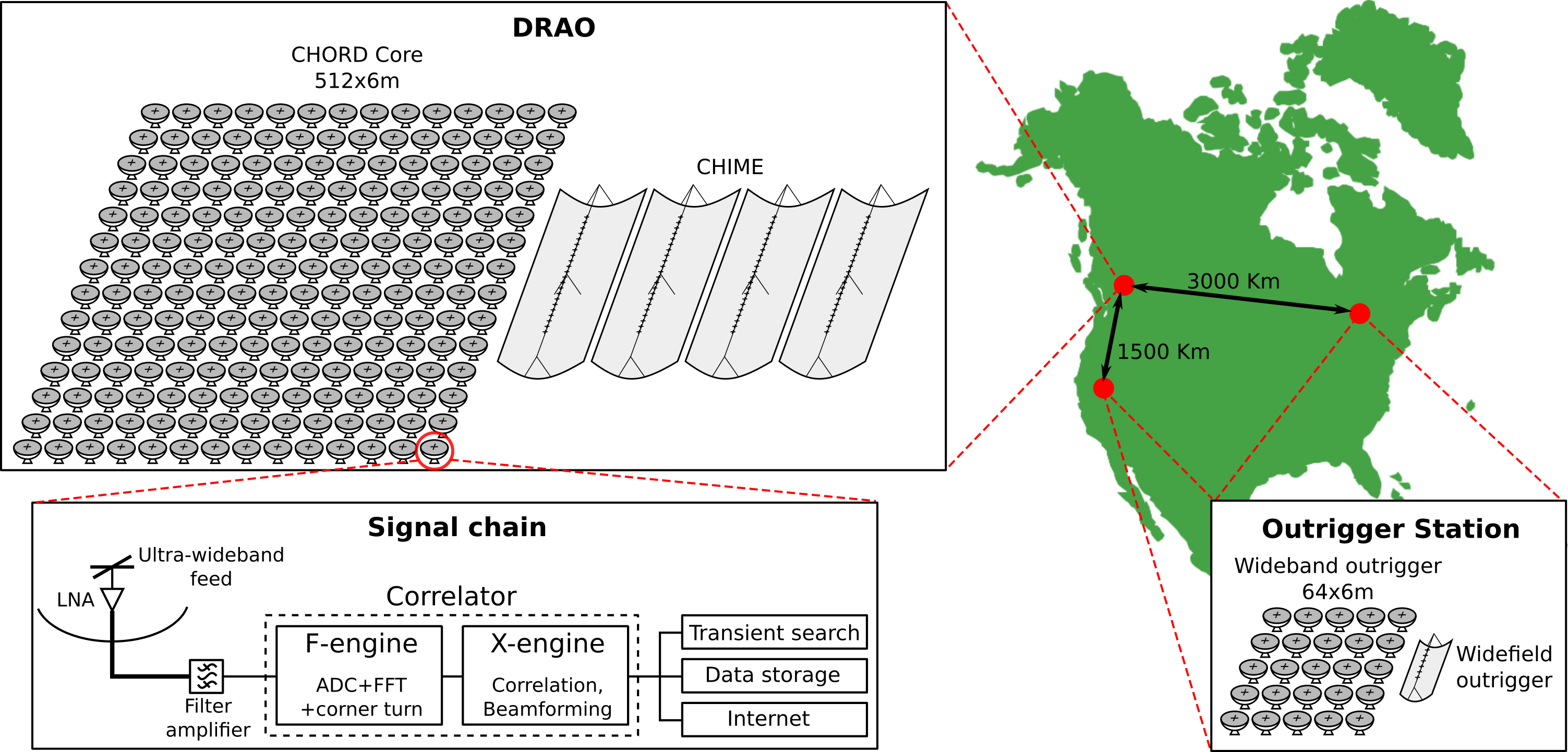}
    \caption{Overview of the CHORD instrument. A large core of UWB 300-1500MHz dishes is located adjacent to CHIME at DRAO. This core is supplemented by a pair of outrigger stations, each equipped with a small CHIME-like 400-800MHz cylinder and an array of UWB 300-1500MHz dishes.
    Signals from the dishes and cylinders are processed by a digital backend which substantially builds on the designs used in CHIME, leveraging new technologies from a telecommunications and artificial intelligence industries.
    {\em Note: Dishes are not drawn to scale, so array images are representative only: for visual simplicity they do not attempt to show the large number of elements.}}
    \label{fig:system_overview}
\end{figure}

The Canadian Hydrogen Observatory and Radio transient Detector (CHORD) is a proposed next-generation radio instrument, designed to build directly on the success of the Canadian Hydrogen Intensity Mapping Experiment (CHIME). CHORD will incorporate CHIME’s best innovations alongside new Canadian technology. Small cylinders derived from the CHIME design and operating from 400-800MHz will be deployed at remote outrigger sites and provide milli-arcsecond-level localization of radio transients. These will be complemented by focused arrays of 6m composite dishes at each site, instrumented with novel ultra-wideband (UWB) feeds, covering a 5:1 radio band from 300–1500MHz. A central array of 512 high-precision 6m composite dishes will anchor the new telescope, providing a huge collecting area with tightly-controlled systematics for breakthrough sensitivity. Figure \ref{fig:system_overview} shows an overview of the system.

%\todo{NEED OVERALL DIAGRAM HERE?}

%The proposed infrastructure represents a substantial leap forward.
%These enhancements are possible thanks a new technique for fabricating low-cost moderate-size radio dishes with precision surfaces, pioneered by the NRC. These composite surfaces, with well understood and controlled systematics, are the building blocks that allow CHORD to make precision measurements, an ideal complement to CHIME’s cylinders, and the foundation for a powerful discovery machine. Newly-developed UWB feeds allow a massively enhanced wide-spectrum view of the sky. Advances in amplifier noise temperature allow for sensitivity to increase by a factor of two.

%This revolutionary new telescope will offer observational capabilities unprecedented in radio astronomy, including higher wideband mapping speed than any other instrument in the world.
CHORD will allow Canadian astronomers to address three of the most exciting areas of astrophysics: (1) elucidating the nature of Fast Radio Bursts (FRBs) and their precise location within galactic hosts; (2) mapping the distribution of matter on cosmic scales to reveal the detailed evolution of structure in the Universe; and (3) measuring fundamental physics parameters, such as probing neutrino properties and testing General Relativity. We envision CHORD as a flagship of Canadian science built firmly on a foundation of Canadian innovation.

Using bandwidth $\times$ collecting area $\times$ sensitivity as a figure of merit, CHORD will be an order of magnitude more powerful than CHIME and be the world-leading facility of its type. This will allow our team to address questions at the frontier of astrophysics and cosmology.

%Canada has established itself as a leader in using astrophysical probes for fundamental physics, from pioneering measurements of the cosmic microwave background fluctuations, to measurements of neutrino oscillations at the Sudbury Neutrino Observatory’s (SNO), which received the 2015 Nobel Prize in Physics.
%CHORD will solidify Canada’s leadership by writing the next chapter of this exciting new era of using radio observations to explore the cosmos.
%Building directly upon CHIME’s success, 

%CHORD will allow our team to amass an unprecedented evidentiary base to address three of the most urgent questions in astrophysics: (1) elucidating the nature of fast radio bursts (FRBs) and using them as a probe of cosmic structure; (2) mapping the distribution of matter on cosmic scales to reveal the physics of cosmic acceleration; and (3) measuring fundamental physics parameters, such as probing the total mass of the neutrinos. 

%CHIME has been revolutionary for its unprecedented data processing capability. It relies on a hybrid correlator, a type of digital processor designed by Vanderlinde and Dobbs. The correlator was created with custom-built electronics and adapted commodity GPUs, and is an order of magnitude less expensive than its predecessors. This is despite having the power to process 13 terabits of data/second—more than all Canada’s internet traffic (CISCO, “Canada 2020” forecast). CHIME is creating a 3D map of the Universe, charting the greatest volume ever surveyed. 

\begin{wrapfigure}{r}{0.4\linewidth}
\vspace{-0.5cm}
    \includegraphics[width=\linewidth]{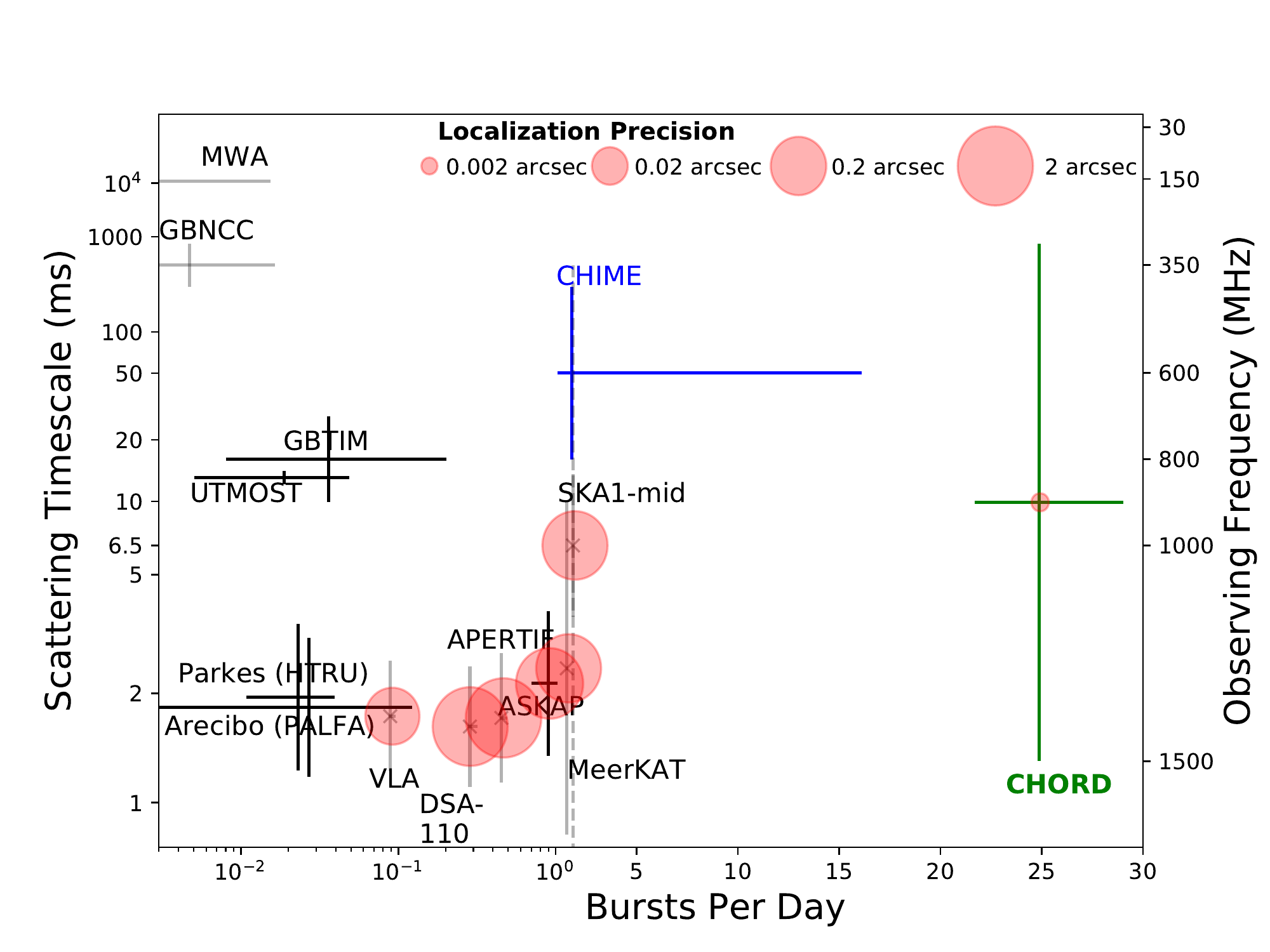}
\vspace{-0.75cm}
    \caption{Present and upcoming FRB surveys. Horizontal extent for each survey shows the range of rates possible given uncertainties in FRB spectral properties, following \citet{2017ApJ...844..140C}. CHORD shows exceptional detection rates (rightward) and frequency coverage (vertical extent), and localization ability (circle size, ~milli-arcsec, compared to ~arcsec on other localizing experiments). Note that the x-axis is logarithmic below one burst per day, to allow distinction among lower-rate surveys.}
\vspace{-1cm}
    \label{fig:frb_rate}
\end{wrapfigure}
\section*{Driving Science}
CHORD will be a breakthrough facility across a wide range of topics. Its design is driven by three primary goals:

\paragraph{Fast Radio Bursts:} FRBs are mysterious, millisecond bursts of radio light of unknown origin coming from far outside our Milky Way galaxy. Though puzzling, they offer new windows on the cosmos and the structure of the Universe \citep{2019ARA&A..57..417C}.
Canada leads the world in FRB discoveries, through the success of CHIME/FRB \citep[e.g. ][]{2019Natur.566..230C, 2019Natur.566..235C}.

One major challenge in understanding FRBs is that most radio telescopes capable of detecting them are poor at determining their location on the sky. Yet FRB localization is key for understanding their nature and for harnessing their potential as cosmic probes \citep{2017ApJ...849..162E}. For an individual FRB, identifying the type and redshift of its host galaxy -- as well as the FRB's location within that galaxy -- can tell us about the source energetics and environment. When studied as a population, such data heavily constrain FRB progenitors.
A localization in the disk of a spiral galaxy suggests youth while an FRB in the outskirts of an elliptical galaxy implies an old progenitor.
A location near a galaxy’s supermassive black hole could imply an interaction with it or its accretion disk \citep{2017MNRAS.471L..92K}.

\begin{wrapfigure}{r}{0.4\linewidth}
\vspace{-2.3cm}
\centering
    \includegraphics[width=\linewidth]{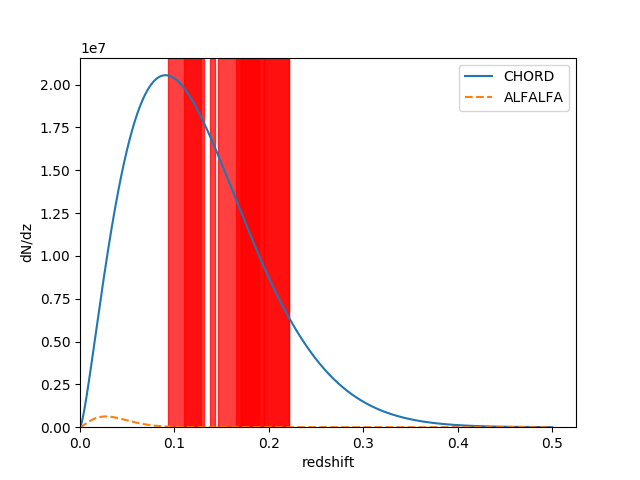}
\vspace{-0.75cm}
    \caption{Predicted dN/dz for CHORD for detections of galaxies in neutral hydrogen, compared to the current state-of-the-art ALFALFA survey.  CHORD represents a factor of 4 increase in peak redshift detection efficiency and an orders-of-magnitude increase in object counts.  Red shaded areas show frequencies where GNSS satellite systems broadcast and RFI removal will be more challenging.}
    \label{fig:gal_survey}
\vspace{-1.0cm}
\end{wrapfigure}

Moreover, knowing their distances enables large scale structure studies and unique determinations of the content of the intergalactic medium, such as constraints on the location of the “missing baryons” in the Universe. Rotation measures of polarized FRBs could constrain the origin and evolution of cosmic magnetism. FRBs present in the early Universe could constrain the reionization history \citep{2016JCAP...05..004F}.

CHORD will provide localizations for the many thousands of FRBs needed for these studies. While repeating FRBs can eventually be localized by following up with other instruments, the large majority of FRBs thus far exhibit only a single burst, offering no such opportunity. CHORD’s outrigger stations overcome this problem, measuring precise coordinates in real time for events detected by CHIME and by the enhanced core. With an unparalleled localization rate (Figure 2), CHORD will revolutionize the field.

\paragraph{Distribution of Matter in the Universe:} CHORD will create the largest 3D image of the cosmos to date, extending CHIME’s all-sky map to finer angular resolution and to cover a huge volume from our cosmic backyard to beyond redshift 3.

CHORD will be exceptionally powerful in carrying out surveys in and around individual galaxies out to cosmological distances via the 21cm emission line. Large single dishes have long been used to measure the atomic gas properties of the local galaxy population, but their small fields of view are a major limitation: ALFALFA at Arecibo, the largest 21cm survey carried out to date, covered less than a quarter of the northern hemisphere sky and detected 30,000 galaxies.
We forecast that CHORD will detect as many as $10^7$ unresolved galaxies across the northern sky, a factor of 300 improvement (Fig \ref{fig:gal_survey}).
This will enable the first truly cosmological studies using atomic gas in individual galaxies via all-sky cosmic flow maps and measures of gas-rich galaxy bias as a function of environment and redshift. In particular, CHORD will be unmatched in its ability characterize the properties of the smallest gas-rich galaxy building blocks and to map accreting circumgalactic medium gas in the local volume well into the SKA era.
%Spekkens comments:
%** That number (10^7) makes me nervous since beyond z~0.1 the galaxies will be confused by the large CHORD beam; hence the "as many as"
%***ALFALFA FOV = 7000deg^2; not sure how you're pitching sky coverage but "less than a quarter" stems from 7000/31000 = 0.23
%**** Not sure how technical you want to get, but Local Volume <= 10ish Mpc.
%***** The 3sigma CHORD sensitivity here is ~10^17 atoms/cm^2, 5x better than anything the SKA precursors will do.
%****** (building blocks) ie. M_HI \lesssim 10^6 Mo.

\begin{wrapfigure}{r}{0.4\linewidth}
\vspace{-0.8cm}
\centering
    \includegraphics[width=\linewidth]{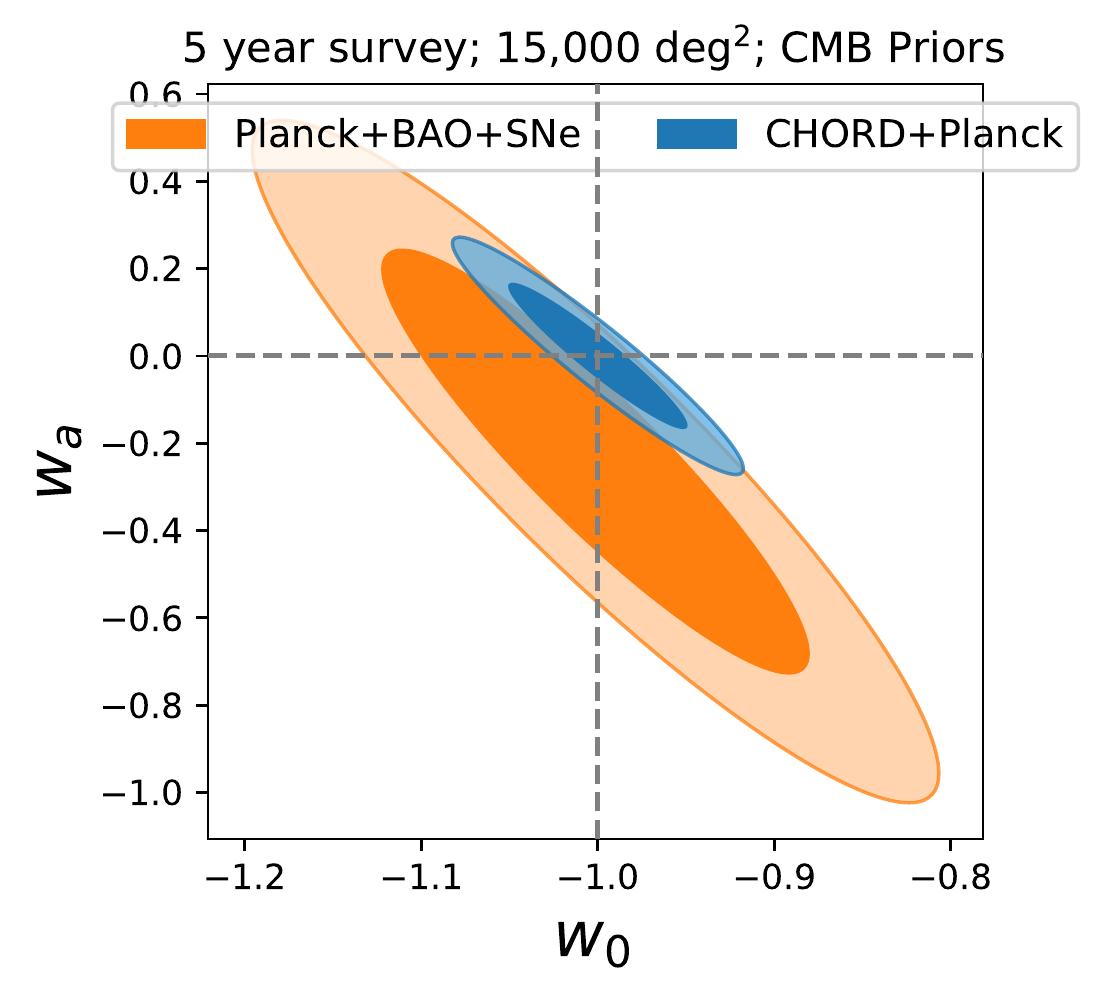}
\vspace{-0.75cm}
    \caption{Forecast constraints in the $w_0,w_a$ plane for CHORD+Planck.  The orange contours are the current state of the art from Planck + current BAO + supernovae while blue are CHORD with Planck priors.  CHORD shrinks the allowed parameter space by nearly a factor of 9 (FoM of 97 vs. 850). With its broad redshift coverage, CHORD is relatively even more sensitive to dark energy models that deviate from the simple $w_0,w_a$ parameterization at high redshifts.}
    \label{fig:w0wa}
\vspace{-0.5cm}
\end{wrapfigure}

Simultaneously, CHORD will transform efforts to map structure in the Universe through intensity mapping, measuring the collective signal from many galaxies. CHORD’s compact dish array is ideally suited to map a billion pixels of the local universe in three dimensions and infer the primordial initial conditions.  Our current best indication of early universe physics is the scalar spectral index $n_s$, where deviations from unity indicate that the universe is not scale invariant.  CHORD will improve on the current best measurements from all of cosmology by a factor of 2.5, from $n_s=0.9665 \pm 0.0038$ \citep{Planck2018params} to an uncertainty of $\pm 0.0015$.  This marks an improvement over even future CMB experiments such as the Simons Observatory
%, which will measure $n_s$ to an error of $\pm 0.002$
\citep{SO2019}.

CHORD will particularly shine in its measurement of the evolution of dark energy. 
The simplest evolving dark energy model has the equation of state $w(a)=w_0+w_a(1-a)$.  As shown in Figure \ref{fig:w0wa}, CHORD will shrink the allowed parameter space in the $w_0,w_a$ plane by more than a factor of eight compared to all current cosmology measurements.  
%This is a hundred times more information than any other survey.
This high statistical precision will substantially improve on CHIME's ability to constrain dark energy (a factor of 4 improvement in the $w_0,w_a$ plane), and is competitive with future dark energy-focused ground and space-based missions in the hundred million to billion-dollar class.
CHORD will have a 20\% smaller area in $w_0,w_a$ than the upcoming Dark Energy Spectroscopic Instrument \citep{DESI2016} even in their most optimistic forecast. %(dark energy figure of merit of 850 for CHORD \textit{c.f.} 704 for DESI).
CHORD and DESI will additionally survey substantially overlapping areas of sky, which will be a powerful tool for mitigating CHORD systematics.
The cross-correlation of CHORD with DESI will have half the overall significance of the CHORD autospectrum enabling sensitive tests of systematics.   
%, and will enable the study of neutrino clustering, which is sensitive to neutrino masses. 

\paragraph{Fundamental Physics:} CHORD will explore new discovery spaces for fundamental physics, including novel tests of relativistic gravity, detection of galaxy-scale gravitational waves, and a probe sensitive to neutrino masses. 

Only 2,800 pulsars are known, a small fraction of the observable Galactic population. CHORD has the potential to dramatically increase this number, with forecasts showing it to be well beyond 10,000. Finding large numbers of pulsars raises the likelihood of discovering relativistic binary neutron-star systems, which are excellent laboratories for testing theories of relativistic gravity. CHORD could discover the first pulsar/black hole binary—long-sought for its potential for stringent and unique relativistic gravity tests. Precision timing measurements of newly-discovered millisecond pulsars can be used as part of an ensemble as a natural telescope to study gravitational waves hypothesized to be produced from merging supermassive black holes. This could offer novel studies of black holes to provide a needed complement to analogous Nobel-class work by LIGO and Virgo. 

Neutrinos are a leading experimental frontier in the standard model of particle physics. Canada has played a central role through the Sudbury Neutrino Observatory (SNO). Such laboratory-based studies measure mass differences between types of neutrinos, but the absolute mass is unknown. One promising approach to measuring the absolute mass is by using cosmic maps to compare the clustering of neutrinos to that of cold dark matter. CHORD’s excellent angular resolution and sensitivity will enable 21cm intensity maps with a thousand-fold increase in information over existing surveys, particularly at small angular scales where neutrino signatures are most prominent. The cosmological measurements that CHORD will provide are urgently needed to inform current intense theoretical work to measure the total mass of these particles in the Universe.

\section*{Enabling Technologies}
\begin{wrapfigure}{r}{0.44\linewidth}
\vspace{-.5cm}
\centering
    \includegraphics[width=0.95\linewidth]{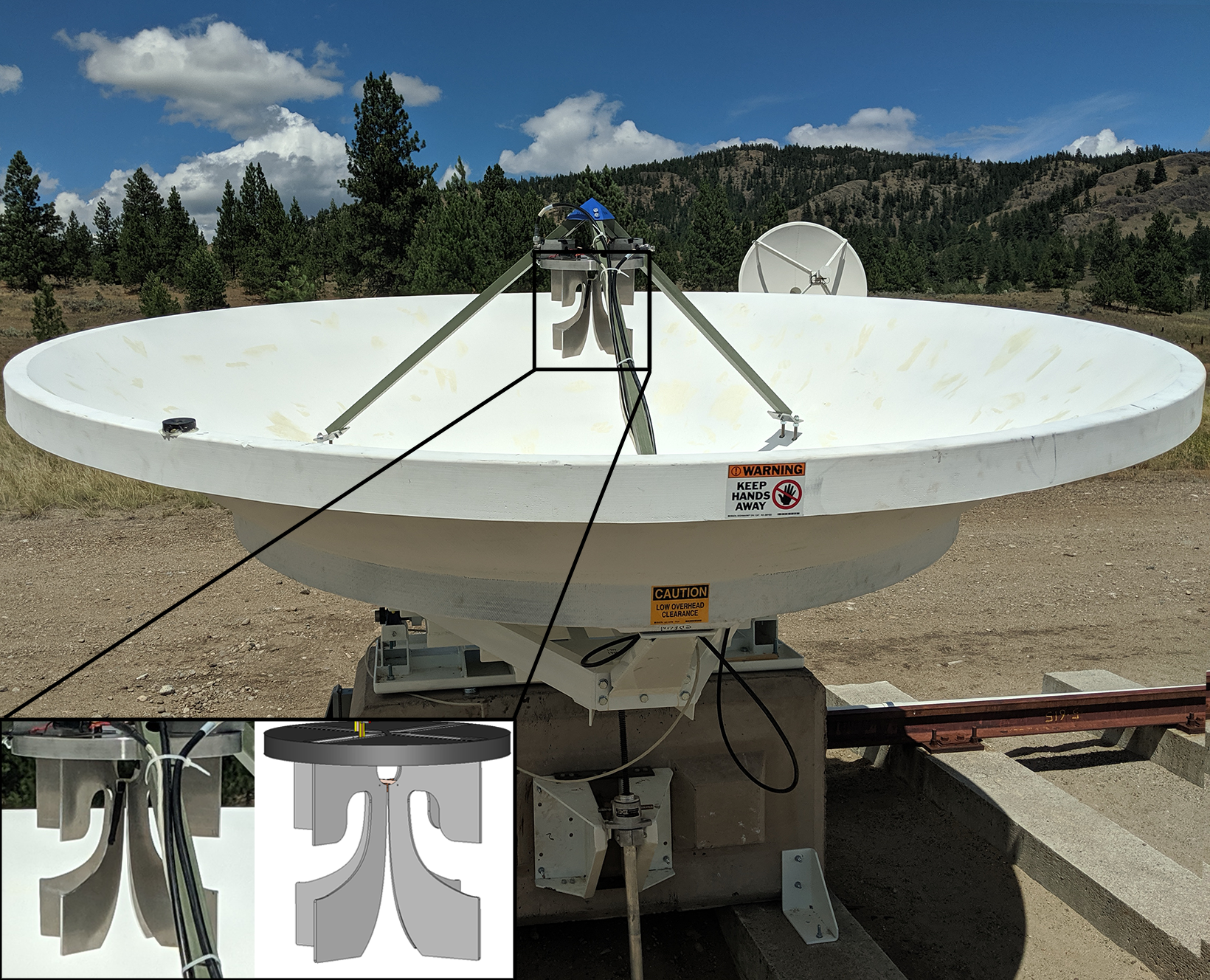}
\vspace{-0.25cm}
    \caption{Prototype composite dish, %produced by NRC, 
    %developed at UToronto 
    deployed at Canada's Dominion Radio Astrophysical Observatory (DRAO).
    A UWB feed is mounted at the focus, with detail shown inset at lower left, alongside CAD model.
    }
    \label{fig:dish_feed}
\vspace{-1cm}
\end{wrapfigure}

A number of critical technologies, pioneered and developed in Canada, have come together to enable CHORD.
The digital innovations that enabled CHIME will be built upon and enhanced, using novel Canadian-driven composite dish and ultra-wideband receiver technologies.

\paragraph{Composite Dishes:}
CHORD's large and precise collecting area is possible at modest cost thanks to a new technique for fabricating low-cost moderate-size radio dishes with precision surfaces, pioneered by our partners at Canada's National Research Council (NRC). These composite surfaces, with well understood and controlled systematics, are the building blocks that allow CHORD to make precision measurements, an ideal complement to CHIME’s cylinders, and the foundation for a powerful discovery machine.
A prototype dish is shown in Figure \ref{fig:dish_feed}.

\paragraph{Ultra-Wideband Feeds:}
The wireless telecommunications industry has invested heavily in exploring novel feed designs. By selectively incorporating and optimizing these, astronomers have developed UWB feeds which can be mass-produced at low cost, and show excellent dish coupling.
Preliminary design work has demonstrated a dual-polarization feed produced from waterjet- or laser-cut sheet aluminum, which achieves outstanding performance over the requisite 5:1 band, in a compact form.
A prototype UWB feed is shown mounted at the focus of the dish in Figure \ref{fig:dish_feed}.

\paragraph{Ultra-low-noise Amplifiers:}
%As for the UWB feeds, receiver circuitry and, in particular, the
Recent developments have enabled room-temperature ultra-low-noise amplifiers \citep[uLNAs; e.g.][]{LNA2}, critical for the large numbers of detectors used on CHORD.
%These are required to operate over  5:1 band, which is significantly wider than than the 3:1 bandwidth of the commercial ultra-wideband systems. As such,
The ultra-wide 5:1 bandwidth necessitates novel design strategies,
as recently demonstrated by Canadian researchers in 2:1-band uLNAs \citep{LNA3}.
%The uLNA design will exploit the transistor-intrinsic gate-drain parasitic capacitance in order to manipulate the input impedance and optimum reflection coefficient of the LNA without incorporating lossy matching components at the LNA input which increase LNA noise. This approach has been shown to be advantages for simultaneous noise and power matching and will be extended to wider bandwidth by employing more sophisticated uLNA loading networks.

\paragraph{Digital Signal Processing:}
Canada is a world leader in radio signal processing systems, as demonstrated in CHIME, which boasts the world's most powerful correlator. It has furthermore amply demonstrated the flexibility and adaptability of such systems, reprocessing the CHIME data in real-time to feed multiple commensal surveys, most notably the CHIME/FRB search engine.

CHORD will build on this foundation, taking advantage of major innovations from recent years. Ever-increasing demands for network speed have given rise to  powerful new data transport technologies. Deep learning applications have driven an explosive growth in low-precision arithmetic performance, continuing Moore's Law with operations equally well suited to radio astronomical processing. These allow an order-of-magnitude increase in the available compute power from recent years, enabling the ultra-wideband signals and compute-intensive Very-Long Baseline Interferometry (VLBI) correlation.

\section*{Telescope Description}
%CHORD will take advantage of advances in digital and radio-frequency technologies to allow Canadian astronomers to study the transient and cosmic-radio sky in ways that were previously not possible. This world-leading next-generation instrument will build on the team's game-changing digital correlator technology developed for the CHIME telescope, alongside new wideband technology and NRC investment in cutting-edge composite reflectors. 

%CHORD is designed to be constructed in phases, to come partially online midway through CHIME’s life and to ultimately fully supersede CHIME. 
CHORD consists of a highly-optimized combination of distributed telescope stations, each adding capabilities to yield a combined instrument with transformative survey and localization abilities. Three elements work together to deliver a host of improvements beyond CHIME, currently the world’s most sensitive FRB detector and fastest mapper of the radio sky. CHORD will deliver a ten-thousand-fold improvement on CHIME’s angular resolution, while providing an order-of-magnitude increase in wideband survey speed. 

\paragraph{Wide-field Outrigger Stations:}
CHORD begins by enhancing the existing CHIME/FRB instrument with two outrigger stations at distances of 30–3,000 km from CHIME.
The huge sensitivity of CHIME mitigates the need for large outrigger stations. At each station, a 30 m-long cylindrical reflector similar to those used on CHIME will be deployed, covering an extremely wide 200-sq degree view of the sky, matched to that of CHIME.
When CHIME detects a cosmic-radio transient signal, the custom digital backend at each outrigger station will be instructed to transmit the locally buffered data to CHIME, so the signals can be correlated together. This synthesizes an instrument with the resolving power of a 1,000 km-wide telescope, allowing unique identification of FRB host galaxies, and the environments within those galaxies where FRBs originate.

The project has captured the attention of international foundations, and seed funding to establish outrigger stations and build the cylinders has already been secured.
%by the Gordon \& Betty Moore Foundation.
%with seed funding from the Gordon \& Betty Moore Foundation.
We have experience working with DRAO to develop and operate telescopes and we are working with them to secure potential sites to host the outriggers.
Candidate sites have been identified and are under study.
%These stations will take two years to be fully commissioned.

\paragraph{Ultra-Wideband Dishes:} To enhance the sensitivity for localization of faint FRBs, each outrigger station will be augmented with an array of 64 six-meter reflectors, coupled to UWB feeds covering a frequency range from 300 MHz–1.5 GHz. These will focus on a common region within the CHIME field-of-view, seeing more deeply over a wider frequency band, providing information about the FRB emission mechanism and the nature and geometry of the intervening plasma. A common set of electronics at each site will run both the dishes and the cylinders, and the data will be combined with CHIME, providing increased sensitivity and angular resolution to the faintest flashes in the sky. 

\paragraph{A Transformative Core:} A large ``core'' array of 512 close-packed 6-m composite reflectors will be deployed adjacent to CHIME on the NRC site at DRAO. The reflectors will be constructed on-site with composite materials developed by NRC and will be instrumented with new, wideband 300 MHz–1.5 GHz receivers. This array provides a many-fold increase in mapping speed over CHIME, providing a hyperspectral data cube unprecedented in radio astronomy, as well as higher wideband survey sensitivity than ever before deployed. It will give a deep, wideband view matching the outrigger dishes, massively increasing the sensitivity to rapid events such as FRBs and pulsar signals, and to the static radio sky. Periodic repointings of this array and the dish outriggers will allow CHORD to tile much of the northern sky. The core will make CHORD a truly transformative instrument for detecting and studying fast radio transients, as an unparalleled cosmic surveyor which sees from present day back through 12 billion years of cosmic history, and as a source of unique radio-sky survey data for the Canadian astronomical community.

\begin{table}[ht]
    \centering
    \begin{tabular}[t]{l|c} \hline
Core            &  \\ \hline \hline
UWB Dishes     & 512  \\
\hspace{12pt}  Diameter          & 6m \\
\hspace{12pt}  Focal ratio       & 0.25 \\
\hspace{12pt}  Radio band        & 300 -- 1500MHz \\
\hspace{12pt}  Field of View     & 130 -- 5 deg$^2$ \\
Collecting Area     & 14,400 m$^2$ \\
Formed Beams   & 512 \\
Beam FWHM    & 17 -- 3 arcmin \\
T$_{sys}$           & 30K \\
SEFD                & 9Jy \\ 
\hline
    \end{tabular}
    \hspace{1em}
    \begin{tabular}[t]{l|c} \hline
%Property            & Specification \\ \hline
Outriggers & \\ \hline \hline

Stations  & 2 \\
Longest baseline    & $\approx 3000$km \\
Burst Localization & $\lesssim$ 10 mas\\
UWB Dishes  & 64 / station\\
\hspace{12pt}Specs & as in Core \\
Cylinders  & 1 / station\\
\hspace{12pt}  Length            & 90m \\
\hspace{12pt}  Width             & 10m \\
\hspace{12pt}  Focal ratio       & 0.25 \\
\hspace{12pt}  Radio band        & 400 -- 800MHz \\
\hline
    \end{tabular}
    \caption{CHORD Technical Properties.}
    \label{tab:my_label}
\end{table}

\section*{Timeline \& Cost}
We are seeking \$22M from the Canada Foundation for Innovation to fully fund the construction and operation of CHORD.
%Substantial matching has already been secured from the Gordon \& Betty Moore Foundation, the Perimeter Institute, and is anticipated from the National Research Council (NRC).
The design and costing are based on our previous experience with the CHIME and CHIME/FRB projects, both of which were delivered on time and on budget and are meeting or exceeding performance specifications. Costing for the dishes has been provided by NRC. Seed funding to design and prototype the outrigger stations has been secured from the Gordon \& Betty Moore Foundation, part of larger in-kind and cash contributions already pledged or secured.
%from partners like the Perimeter Institute and the NationalResearch Council.
CHORD will have very low operations costs: similar to CHIME, which is maintained by a single on-site telescope operator, and reflectors have no motorized parts that can wear out.

The outrigger stations and cylinders are under development already, and will be fully commissioned and operating in 2 years. It will take 1-2 years to finalize the design of the  UWB composite dishes of the core and the outrigger dishes, and 2-3 years to fully construct once funding is secured.

All technologies have been demonstrated in the laboratory or in the field. Unit and Integration testing, scaling, and optimizing performance constitutes a substantial development effort. Eleven engineer-years of effort spread is budgeted for firmware, RF, and mechanical engineering.

%Software data pipeline development: CHORD is a digital instrument, which generates and processes enormous volumes of data, comparable to the total traffic on the Canadian Internet. Handling and processing these data to yield the required data products requires sophisticated algorithms and detailed engineering. We budget 20 person-years of effort (5 programmers over 4 years) to develop and commission the required software pipelines so that the data can be made available in near real time.

%There is no similar infrastructure available anywhere on Earth—this new instrument is only now possible to build because of new technology innovations. We will work closely with CIRADA, a CFI-IF national initiative led by members of our team that was funded to develop software and algorithmic infrastructure for converting the extreme data rates and complex multi-dimensional outputs of modern radio telescopes into accessible, science-ready, advanced data products. The data challenges and science goals of this present proposal are consistent with those of CIRADA and complementary to Canada’s long-term plan for SKA. We will thus leverage those efforts. 

%CHORD’s reflectors will work in a drift scan mode, continuously observing the sky as the earth rotates, which maximizes its potential by mapping the cosmos and searching for radio transient events 24 hours a day, 7 days a week. CHORD will experience virtually no downtime. 

\section*{National and International Context}
CHORD will be internationally unique. By a wide margin, it will be the pre-eminent telescope on Canadian soil and thus a worthy successor to CHIME. Internationally, CHORD occupies a critical position within the landscape of current- and next-generation instruments.

The proposed HIRAX \citep{2016SPIE.9906E..5XN} and DSA-2000 \citep{2019arXiv190707648H} projects each bear some similarity to CHORD, being composed of large numbers of small dishes, but differ in many crucial respects. HIRAX is in the southern hemisphere and observes a narrower 400–800 MHz band -- in many ways it is a sister project to CHORD, with some overlapping personnel and shared R\&D. DSA-2000 will observe similar frequencies to CHORD, but with dishes distributed over thousands of square kilometers. CHORD’s dense core, broad bandwidth, and interface with CHIME uniquely enable its distinct science portfolio. 

Optical galaxy surveys such as DESI provide an independent probe of cosmic structure that is complementary to 21 cm intensity mapping. For science as fundamental as dark energy, it is essential to follow multiple lines of inquiry. CHORD is a low-cost avenue to this physics, able to probe additional modes, and subject to different systematics. Cross-correlation between surveys will provide results robust against contaminating systematics: CHORD is a critical ingredient in the global drive to quantify dark energy physics.

The SKA is a multi-billion-dollar international effort, scheduled to come online midway through the next decade. As a world observatory, it will address a broad spectrum of questions at the forefront of radio astronomy. As a specialized experiment, CHORD is optimized for rapid transient sensitivity and large-scale cosmic mapping, rather than as a broad observatory. It will complement and enhance the science reach of the SKA, strengthen Canadian astronomers for leadership roles, develop Canadian technology (particularly data handling and signal processing algorithms), and put Canadians at a unique advantage globally in the SKA era.

\section*{Membership \& Data Sharing}

CHORD is a pan-Canadian effort. We expect it to deepen existing collaborations and to help the team forge new ones to keep Canada at the forefront of global astronomy. The instrument is also designed to nurture multidisciplinary research as it will help us address questions that are of equal interests to physicists as well as astronomers. Interpreting CHORD’s results will also require a range of research strategies, including mathematical and physical theory, not to mention large and complex computer simulations.
%CHORD will thus create a vast space for further innovation.

CHORD's survey will generate a rich data set, which will be made rapidly available to the wider Canadian astronomical community.
Data products, catalogues and tools associated with CHORD will be publicly released in order to amplify and broaden the science output from CHORD, and to create opportunities for scientists to make other discoveries from our data. All public releases will first need to pass a quality assurance process, and some releases may be subject to a proprietary period before being made public, to provide scientific opportunities for HQP working directly on CHORD. Subjecting a data set to quality assurance, preparing it for public release, and then providing access to it in perpetuity are all significant undertakings, which go far beyond simply “writing up the result” as pursued for typical isolated observations or experiments. We will thus work closely on all these tasks with CIRADA and the CADC, who are developing considerable experience on similar public releases for VLASS, ASKAP and CHIME. 

\section*{Criteria}

% ****BEGIN CRITERIA SECTION (4 page limit) ****

%Instructions:
%Assessment and prioritisation of facilities and programs in LRP2020 will be based on a predefined set of criteria. 
%Authors are requested to explicitly address these criteria in the set of text boxes below. Some criteria may not be applicable to all white papers. 
% IMPORTANT: 
% There is no specific length limit on individual boxes. 
% However, the full set of 8 boxes should comprise no more than 4 pages and these pages **do** count toward the 10-page limit of the full document.

\begin{lrptextbox}[How does the proposed initiative result in fundamental or transformational advances in our understanding of the Universe?]
%insert your text here
CHORD will take advantage of advances in digital and radio-frequency technologies to allow Canadian astronomers to study the transient and cosmic-radio sky in ways that were previously not possible. This world-leading, next-generation instrument will build on the team's game-changing digital correlator technology developed for the CHIME telescope, alongside new wideband technology and NRC investment in cutting-edge composite reflectors. 

\vspace{-0.3cm}
\paragraph{Fast Radio Bursts:}For an individual FRB, identifying the type and redshift of its host galaxy—as well as the FRB's location within that galaxy—unlocks volumes about the source energetics and environments. While repeating FRBs can eventually be localized by following up with other instruments, the large majority of FRBs thus far exhibit only a single burst, offering no such opportunity. CHORD's outrigger stations overcome this problem, measuring precise coordinates in real time for events detected by CHIME and by the enhanced core. With an unparalleled localization rate, CHORD will revolutionize the field.

\vspace{-0.3cm}
\paragraph{Distribution of Matter in the Universe:}CHORD will create the largest 3D image of the cosmos to date, extending CHIME's all-sky map to finer angular resolution and to cover the full volume from our cosmic backyard to beyond redshift 3. This high statistical precision will substantially improve on CHIME's ability to constrain dark energy, enabling a complementary probe to billion-dollar class projects, and will enable the study of neutrino clustering, which is sensitive to neutrino masses.

\vspace{-0.3cm}
\paragraph{Fundamental Physics Parameters:}CHORD will explore new discovery spaces for fundamental physics, including novel tests of relativistic gravity, detection of parsec-scale gravitational waves, and a probe sensitive to neutrino masses. The cosmological measurements that CHORD will provide are urgently needed to inform current intense theoretical work to measure the total mass of these particles in the Universe.
\end{lrptextbox}

\begin{lrptextbox}[What are the main scientific risks and how will they be mitigated?]
%insert your text here
CHORD's ability to map large-scale structure via HI emission depends on our ability to filter this signal from the much brighter foreground radiation emanating from our own galaxy. This separation in turn depends critically on our ability to understand and control the instrumental response. The antennas that make up CHORD's core have been designed from the ground up to minimize these systematics, with dish surfaces reproducible to sub-millimetre accuracy, deep focal ratios to shield cross-talk between dishes, and feeds and support systems mass-produced with extreme tolerances.

CHORD will localize large numbers of FRBs, and many of the studies enabled by these localization will require knowing FRB redshifts, which in turn require host galaxy identification and follow-up. We have a strong record of establishing collaborations and securing time for follow-up, e.g., an extensive campaign to obtain redshifts on 100s of galaxy clusters detected by the South Pole Telescope (SPT).

Timeliness is critical for CHORD to realize its ambitious goals. FRB science is a competitive and rapidly moving field, and by the time CHORD is built, some questions may have been addressed by others: already multiple facilities have begun announcing localizations.
However, none of these can match or even approach the rates CHORD will achieve. If the first dozen events to be localized come from other instruments, the first thousand will come from CHORD. These large numbers of events open the door to many of the most promising avenues of investigation.
%\kv{timeline -- others may beat us to the punch w/ the first 10, but w/ thousands, things open up.}

CHORD is opening a new phase space in sensitivity / bandwidth / localization capabilities, and in doing so enables a number of speculative science targets -- from finding a pulsar-black hole binary, to detecting strongly-lensed high-z FRBs. Any number of these may fail to pan out, but the space opened by these new capabilities is likely to lead to additional unexpected and high-value discoveries, further ensuring the scientific return of this breakthrough facility.
\end{lrptextbox}

\begin{lrptextbox}[Is there the expectation of and capacity for Canadian scientific, technical or strategic leadership?] 
%insert your text here
Canadian researchers are leading every aspect of CHORD, and will be developing the critical technologies, analyzing the world-leading data set, and making the crucial discoveries.
CHORD will be the most powerful survey telescope over a wide range of radio frequencies for the foreseeable future.
Young researchers working on CHORD will be ideally situated to take international leadership roles, as many of those who worked on CHIME have already done.
%Indeed, CHORD is naturally placed to build substantially on CHIME's successes, which have already put Canadian radio astronomy on the front pages of newspapers, popular media, and scientific journals, and propelled Canadian researches to the top levels of international scientific circles.

CHORD will be a flagship for Canadian Science. CHIME is already delivering spectacularly on ground-breaking science, and CHORD will enable a new level of results. It will allow localizations a thousand times more precise than CHIME; cover a vastly greater spectral coverage of 0.3–1.5 GHz; and increase wideband survey speed by an order of magnitude. 

While CHIME has been revolutionary, much of its underlying technology is already being superseded, largely due to ongoing Canadian ingenuity. CHORD will allow Canada to maintain both the theoretical and experimental leadership CHIME has bought us, and by coming online in a staged way, will deliver the next boost to our reputation in short order. As such, CHORD will allow the Canadian astronomy community to significantly advance a number of urgent fronts of research, and consequently shape the very questions we are asking in the decades to come. 

%CHIME has been revolutionary for its unprecedented data processing capability. It relies on a hybrid correlator, a type of digital processor, designed by Vanderlinde and Dobbs. CHORD will incorporate CHIME's best innovations and new Canadian technology to offer observational capabilities unprecedented in radio astronomy, including higher wideband mapping speed than any other instrument in the world. We envision CHORD as a flagship of Canadian science built firmly on a foundation of Canadian innovation.

%CHORD's ultra-wide 5:1 bandwidth feed will enable observations with a spectrum three times wider than CHIME's. Small cylinders derived from the CHIME design, complemented by focused dish arrays, will be deployed at remote sites and provide sub-arcsecond localization of radio transients. The surface area will be substantially increased by deploying an array of 512 composite radio dishes. This is possible because of a new technique for fabricating low-cost moderate-size radio dishes with precision surfaces, pioneered by our partners at NRC. These composite surfaces, with well understood and controlled systematics, are the building blocks that allow CHORD to make precision measurements, an ideal complement to CHIME's cylinders, and the foundation for a powerful discovery machine. Lastly, advances in amplifier noise temperature will allow for sensitivity to increase by a factor of two. Using bandwidth $\times$ collecting area $\times$ sensitivity as a figure of merit, CHORD will be an order of magnitude more powerful than CHIME and be the world-leading facility of its type. 
\end{lrptextbox}

\begin{lrptextbox}[Is there support from, involvement from, and coordination within the relevant Canadian community and more broadly?] 
%insert your text here
CHORD is a pan-Canadian effort and a progression from CHIME, whose core members have been working together since 2008. The main institutions behind CHORD are UofT, McGill, U Calgary, the Perimeter Institute, and NRC Herzberg, with UBC also playing a role on much of the science.

NRC'S DRAO site is providing on-site infrastructure including laboratories, fabrication shops, a computer network/internet access, office space and power. The DRAO site is protected against RFI by government regulation. In addition to hosting CHIME and the CHORD compact array, DRAO is providing personnel and will assist with environmental assessments and managing construction. 

The team will continue working with its industry partners, including AMD, IBM, Intel and CoolIT. The technological demands of CHORD pose rich challenges for these partners and have already led to several innovations that have been adopted into their products. Several of the team's graduate students and postdoctoral fellows are already working closely with these firms to develop algorithms and testing new data transport protocols and product designs. We will also rely on our long-standing relationship with Compute Canada to store and transmit the enormous amount of data produced by CHORD's correlator, and also the CADC for disseminating the data sets through its public portal.

The instrument is also designed to nurture multidisciplinary research as it will help us address questions that are of equal interests to physicists as well as astronomers. Interpreting CHORD's results will also use of a range of research strategies, including mathematical and physical theory, not to mention large and complex computer simulations. CHORD will thus create a vast space for further innovation. 
\end{lrptextbox}

\begin{lrptextbox}[Will this program position Canadian astronomy for future opportunities and returns in 2020-2030 or beyond 2030?] 
%insert your text here
CHORD will come online in phases to deliver the next boost to our reputation in short order. As such, CHORD will allow the Canadian astronomy community to significantly advance a number of urgent fronts of research, and consequently shape the very questions we are asking through the next decade and beyond.

Internationally, CHORD is already drawing the interest of additional collaborators beyond our partnerships through CHIME. Notably, CHORD's scientific and technical achievements will strengthen Canada's position within SKA, which will come online midway through the next decade. Innovations likely to be transferred to SKA include fast signal processing, GPU-based data analysis and algorithm development, and real-time processing of massive data sets (e.g., Pen sits on SKA's central signal processing board). We are also in talks with the Deep Synoptic Array, the Zwicky Transient Facility and LIGO about sharing technology and science results. Vanderlinde is developing the GPU-based correlator for the HIRAX telescope in South Africa, and Pen is helping enable science on the 500m FAST telescope in China.

CHORD will also benefit the NRC and enhance the capabilities of the DRAO site, providing the infrastructure to serve the broad Canadian astronomy community. 

\end{lrptextbox}

\begin{lrptextbox}[In what ways is the cost-benefit ratio, including existing investments and future operating costs, favourable?] 
%insert your text here
CHORD will be constructed in phases to directly leverage and integrate with existing CHIME infrastructure while delivering near-term, progressive upgrades. The upgraded CHIME will thus remain state of the art as it is transformed into CHORD.

CHORD's reflectors will work in a drift scan mode, continuously observing the sky as the earth rotates, which maximizes its potential by mapping the cosmos and searching for radio transient events 24 hours a day, 7 days a week. Once constructed, the operation and maintenance needs of CHORD are relatively simple given that it has no motorized parts. This means the most time-consuming tasks for the operations of a traditional telescope are unnecessary (i.e., alignment, repair of bearings and motors, allocation of time and scheduling of pointed observations), and the operator tasks and skillsets are similar to those required for maintaining a computer cluster. Once fully operational, the annual operation budget will be \$575,000/year. (Operating costs prior to deployment of the core will be a modest increment to CHIME/FRB operating costs, on the order of \$20K/yr for electricity.) 

CHORD is conceived as a five-year experiment. However, with proper maintenance, it could function for 20 years or more. Should it become clear the facility has potential for ongoing breakthrough science, we will seek long-term operating funds through, for example, the Major Science Initiatives program or NRC. 

\end{lrptextbox}

\begin{lrptextbox}[What are the main programmatic risks
%Instructions: Programmatic risks include but are not limited to schedule, feasibility, budget, technical readiness level, computational or software requirements, dependence on other partners, and governance plan.
and how will they be mitigated?] 
%insert your text here
CHORD is not yet fully funded: seed funding from the Moore foundation is allowing the initial stages of outrigger site preparation and cylinder deployment to proceed, but the timely execution of the full CHORD research programme will depend on securing a full funding package. A joint application is being submitted to the Canada Foundation for Innovation (CFI)'s 2020 Innovation Fund competition.

All technologies required for CHORD have been demonstrated in laboratory settings, and the majority are already operating on-sky in fielded prototype systems. Concerns about handling and processing the deluge of raw data have been largely retired with CHIME's success.

%\kv{Need some text re: politics of outrigger sites. Securing land, respecting rights, etc.}
An open risk is securing sites for the placement of the outriggers. The team are in discussions with existing radio observatories such as Algonquin in Ontario and Green Bank, Owens Valley and Hat Creek in the US. Co-locating at existing observatories allows our team access to radio quiet zones, to leverage existing local knowledge, to navigate environmental and native concerns and to mitigate risks during operations. New communications technologies will continue to degrade the Radio Frequency Interference (RFI) environment, and active engagement will be required to optimize sensitivity. Placement of the site closest to DRAO in BC must consider the terrain, agricultural, environmental and native regulations. The team at DRAO are assisting with finding a suitable site and have a long and successful history of working with local stakeholders.

\end{lrptextbox}

\begin{lrptextbox}[Does the proposed initiative offer specific tangible benefits to Canadians, including but not limited to interdisciplinary research, industry opportunities, HQP training,
%HQP=Highly qualified personnel, defined as individuals with university degrees at the bachelors' level and above.
EDI,
%EDI = equity, diversity and inclusion 
outreach or education?] 
%insert your text here
CHIME and related investments have put Canadian astrophysics centre stage and have thereby raised the profile and reputation of Canadian science. Several Canadian companies and subsidiaries (e.g., IBM, AMD, SkyWatch, CoolIT, Greyback Construction, EDS) have collaborated with us on common interests, including signal processing, large-scale computing, and digital analysis, and then used their success to expand market share. For example, Greyback was awarded the 2014 SICA Commercial Building Award for their work constructing CHIME, while EDS estimates their experience in building major telescopes has spurred over \$400M in other contracts to build theme park rides, bridges, and other projects. 

An important benefit of this project is the human capital it creates, in the form of highly trained individuals that will go on to drive Canada's innovation sectors and create new technology and knowledge. All of the technology development and scientific analysis relies on trainees including students and postdocs. The end-to-end process of designing, building, and commissioning a cutting-edge scientific instrument, then using it for astronomical measurements, provides an ideal training ground for HQP. In addition to academic careers, these trainees will be well positioned to advance Canada's technology interests in telecommunications, time series analysis, and handling of massive data streams. Our former trainees have found success in medical physics, finance, and information technology. One of the firmware programmers who worked on CHIME labs has launched a spin-off designing firmware for scientific applications.

CHORD also presents an opportunity to advance the inclusivity of Canadian astronomy. In composing our research team, we aimed to reflect a breadth of experience and expertise. In addition to a track record of research excellence, this included team members at different career stages, and members with a diversity of backgrounds. The CHORD team is drafting an EDI plan, including strategies to work with diverse collaborators and to recruit diverse HQP. In this respect, we benefit from the expertise of team member Gaensler, who has helped shape EDI policies at Toronto as a member of the University's Equity and Diversity in Research and Innovation (EDRI) Working Group, and through policies he has enacted at the Dunlap Institute as its director. 
\end{lrptextbox}

\setlength{\bibsep}{0pt plus 0.3ex}
\begin{multicols}{2}
\footnotesize{
\bibliography{whitepaper}
}
\end{multicols}

\end{document}